\input{aipcheck}

\documentclass[
    ,final            
  ]
  {aipproc}

\layoutstyle{6x9}

\begin{document}

\title{Towards an optimized design \\ for the Cherenkov Telescope Array}



\author{V. Stamatescu}{
  address={\mbox{IFAE Barcelona,}}
}

\author{Y. Becherini}{
  address={\mbox{APC Paris,}}
  ,altaddress={\mbox{LLR Paris,}}
}

\author{K. Bernl\"{o}hr}{
  address={\mbox{MPI-K Heidelberg,}}
}

\author{E. Carmona}{
  address={\mbox{CIEMAT Madrid,}}
}

\author{P. Colin}{
  address={\mbox{MPI-P Munich,}}
}

\author{C. Farnier}{
  address={\mbox{Stockholm University,}}
}

\author{L. Gerard}{
  address={\mbox{DESY Zeuthen,}}
}

\author{J. A. Hinton}{
  address={\mbox{Univ. of Leicester,}}
}

\author{B. Kh\'{e}lifi}{
  address={\mbox{LLR Paris,}}
}

\author{N. Komin}{
  address={\mbox{LAPP Annecy,}}
}

\author{G. Lamanna}{
  address={\mbox{LAPP Annecy,}}
}

\author{J.-P. Lenain}{
  address={\mbox{Landessternwarte Heidelberg,}}
}

\author{G. Maier}{
  address={\mbox{DESY Zeuthen,}}
}

\author{A. Moralejo}{
  address={\mbox{IFAE Barcelona,}}
}

\author{C. L. Naumann}{
  address={\mbox{LPNHE CNRS-IN2P3 Paris,}}
}

\author{R. D. Parsons}{
  address={\mbox{MPI-K Heidelberg,}}
}

\author{F. Di Pierro}{
  address={\mbox{INAF Torino}}
}

\author{H. Prokoph}{
  address={\mbox{DESY Zeuthen,}}
}

\author{S. Vorobiov}{
  address={\mbox{DESY Zeuthen,}}
}

\renewcommand\XFMtitleblock{
  \XFMtitle
  \let\XFMoldpar\par
  \def\par{\XFMoldpar\def\par{\space
        \mbox{for the CTA Consortium}\XFMoldpar}}
   \XFMauthors
   \let\par\XFMoldpar
   \XFMaddresses
   \XFMabstract
   \vspace{5pt}
   \XFMkeywords
   \XFMclassification
}

\begin{abstract}
  The Cherenkov Telescope Array (CTA) is a future instrument for very-high-energy (VHE) 
  gamma-ray astronomy that is expected to deliver
  an order of magnitude improvement in sensitivity over existing instruments.
  In order to meet the physics goals of CTA in a cost-effective way,
  Monte Carlo simulations of the telescope array are used in its design.
  Specifically, we simulate large arrays comprising numerous large-size, medium-size and 
  small-size telescopes whose configuration parameters are chosen based on current 
  technical design studies and understanding of the costs involved. Subset candidate arrays 
  with various layout configurations are then selected and evaluated in terms of key 
  performance parameters, such as the sensitivity. This is carried out using a number of data 
  analysis methods, some of which were developed within the field and extended to CTA, 
  while others were developed specifically for this purpose. We outline
  some key results from recent studies that illustrate our approach
  to the optimization of the CTA design.
\end{abstract}

\maketitle


\section{Performance of the Candidate Arrays}

\begin{figure}
  \includegraphics[height=.23\textheight]{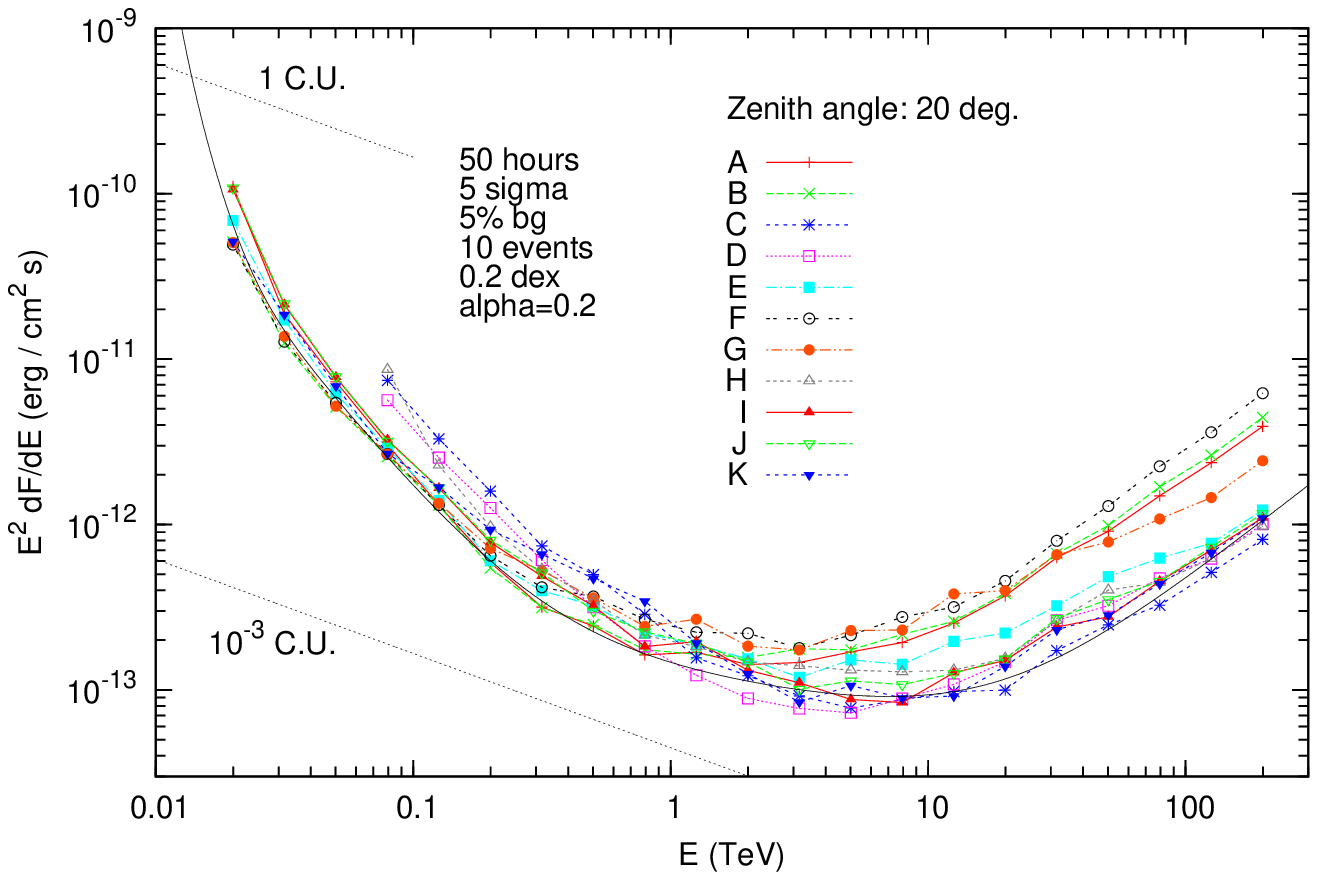}
  \includegraphics[height=.22\textheight]{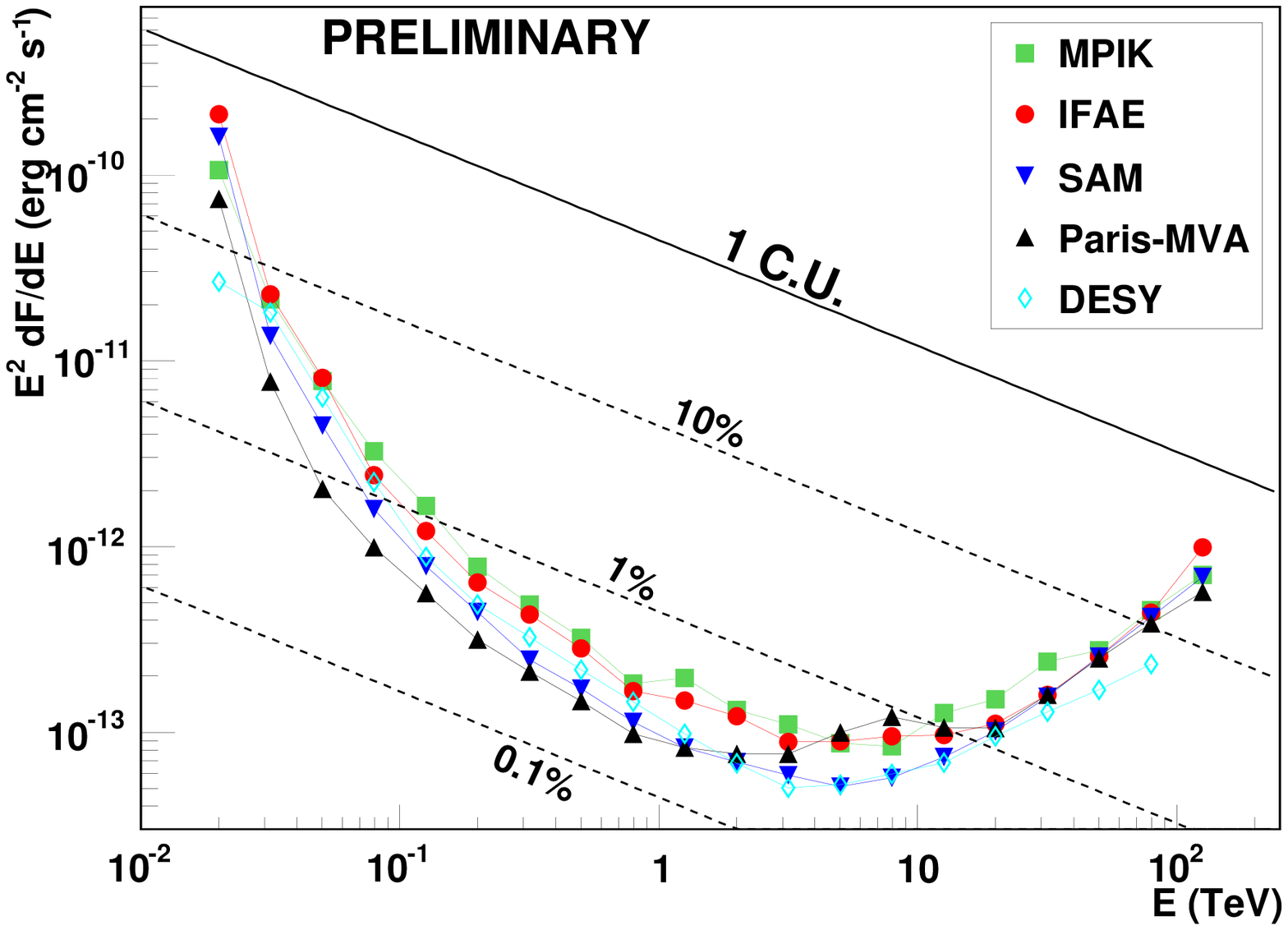}
  \caption{(\emph{Left}) Point-like source sensitivity of 11 CTA candidate array layouts of identical estimated costs 
    for 50 hours observation time and zenith angle of $20^\circ$, evaluated with the baseline analysis method.
    The solid black line approximates the best performance of any of these arrays at any energy,
    excluding D, which is optimized for energies of a few TeV.
    The Crab Unit (C.U.) and milli-C.U. fluxes (see text) are shown for comparison.
    (\emph{Right})
    Sensitivity of balanced array I, evaluated using the baseline/MPIK (green squares)
    IFAE (red circles), SAM (blue triangles), Paris-MVA (black triangles) and DESY (light-blue diamonds) analyses.
  }
  \label{fig1}
\end{figure}

We evaluate the performance of CTA candidate arrays,
which are compared in \mbox{Fig. \ref{fig1}} in terms of their differential
sensitivity, calculated using our baseline analysis \cite{Bernlohr2012}.
The candidate array layouts comprise varying numbers of
large-size (LSTs), mid-size (MSTs) and small-size (SSTs) telescopes that are selected
according to a fixed estimated total cost from a large $275-$telescope configuration known as production 1 (prod$-1$).
Part of the simulations were performed by using a distributed approach within the European Grid Initiative \cite{Komin2011}. 
For the upcoming prod$-2$, a Grid-based simulation and analysis pipeline is being prepared.
Arrays E and I are so-called balanced arrays (balanced in terms of the distribution resources across energy range of CTA),
and thus are well-suited for evaluating the effects of different analysis methods or telescope designs.

We verify that configurations selected to be optimal using the 
baseline analysis also reach close to the optimal sensitivity
when employing alternative analysis methods (see descriptions in \cite{Bernlohr2012,Becherini2012}).
\mbox{Fig. \ref{fig1}} illustrates the effect of these methods on sensitivity for the array I,
which consists of 3 LSTs each with $4.9^\circ$ field of view (FoV),
18 MSTs with $8^\circ$ FoV, and 56 SSTs with $9^\circ$ FoV.
While most methods are under development,
some already approach differential sensitivity levels of 2 milli-C.U.
\footnote{ $\textrm{ 1 C.U.} = 2.79 \cdot 10^{-7} \textrm{m}^{-1} \textrm{s}^{-1} \textrm{TeV}^{-1} \times (E/TeV)^{-2.57}$}
(Crab Unit) in the core range around 1 TeV.
This minimum detectable flux is determined conservatively,
using $5$ bins per decade in energy and demanding, for a given bin,
a minimum $5\sigma$ detection (using Eq. 17 from  Li \& Ma \cite{LiMa1983} and 5 background control regions),
$\geq 10$ $\gamma$-ray events, and that the $\gamma$-ray excess be greater than $5\%$ of the residual background.

\begin{figure}
  \includegraphics[height=.2\textheight]{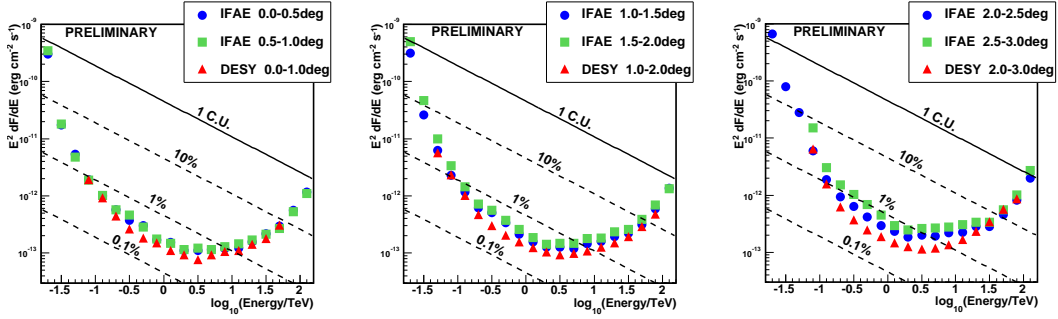}
  \caption{Offset point-like source 50 hour sensitivity of balanced array E,
    for source offsets of between $0$-1$^\circ$ (left), $1$-2$^\circ$ (middle) and  $2-3^\circ$ (right),
    evaluated using two alternative analysis methods: IFAE (blue circles and green squares) and DESY (red triangles).}
  \label{fig2}
\end{figure}

We are also investigating the off-axis performance of CTA.
\mbox{Fig. \ref{fig2}} shows the off-axis differential sensitivity for offset angles out to $3^\circ$
calculated using two analyses for array E, which comprises 4 LSTs with $4.6^\circ$ FoV,
23 MSTs with $8^\circ$ FoV, and 32 SSTs with $10^\circ$ FoV.
While the off-axis analysis algorithms are expected to have room for improvement, these results indicate
that the point-source differential sensitivity in the energy range of $\sim100$ GeV to $\sim10$ TeV
may be well below $0.1$ C.U. out to offsets of at least $\sim3^\circ$ off-axis.

\section{LST optics and camera design}

\begin{figure}
  \includegraphics[height=.22\textheight]{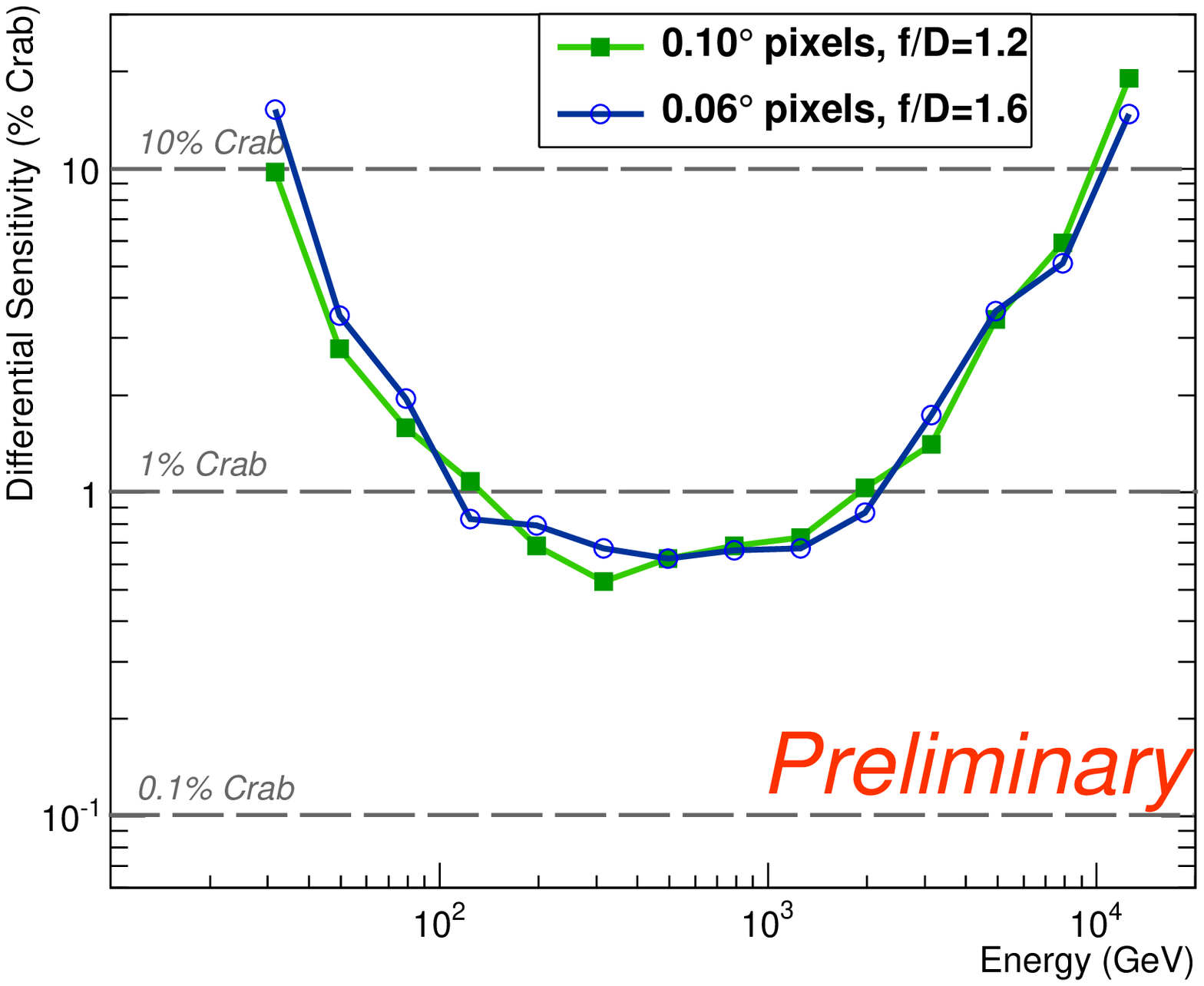}
  \includegraphics[height=.22\textheight]{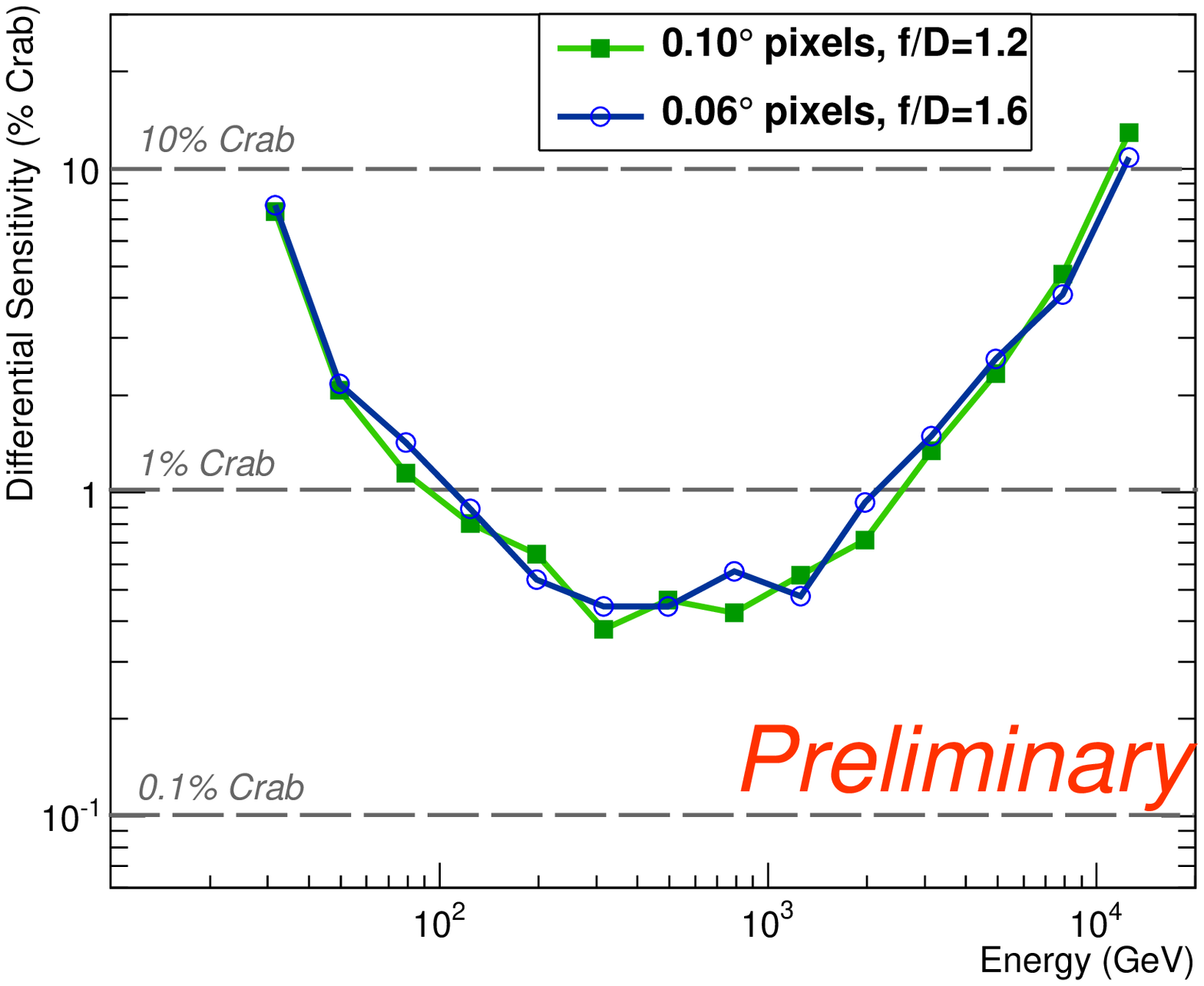}
  \caption{Differential sensitivity of (\emph{left}) 3 LSTs arranged
    in an equilateral triangle and (\emph{right}) 4 LSTs in a square, both of side length of 105 m,
    for realistic (green squares) and ideal (blue open circles) LST designs (see text)
    of the camera pixel size and the telescope optics. Note that the contribution to the 
    background from electrons has not been taken into account in this LST study.}
  \label{fig3}
\end{figure}

The LSTs are simulated in prod$-1$ with a parabolic dish shape
of focal length $f = 31.2 \textrm{m}$ and area $A = 412 \textrm{m}^2$, coupled with $0.09^\circ$ pixels.
Mechanical constraints of the envisaged LST structure impose a maximum $f = 28 \textrm{m}$.
In a dedicated study, we simulate an alternative design with  $A = 382 \textrm{m}^2$ and $0.1^\circ$ pixels,
and use an intermediate dish shape to mitigate the degradation of the optical PSF
given by the reduced $f/D$, where $D$ is the mirror diameter.
We also simulate an idealized design with $A = 382 \textrm{m}^2$, $f = 37 \textrm{m}$ and $0.06^\circ$ pixels.
Results show that the mean Hillas width \cite{Hillas1985} of small gamma-ray images,
with Hillas sizes of around 100 p.e., are 20\% smaller for the idealized case.
However, this does not translate to an improvement in gamma-hadron separation power, as indicated in \mbox{Fig. \ref{fig3}},
which compares the realistic and idealized designs of two LST arrays.

The LST $f/D$ and FoV parameters are being tuned for prod$-2$ by using a toy model,
which estimates the relative sensitivity in the LST-dominated energy range of $30–-100$ GeV, at a fixed LST array cost. 
The model is based on prod$-1$ simulations and assumes that
the background control region size depends on the angular distance between the source position and camera center,
and that sensitivity scales as $N^{-0.5}$, where $N$ is the number of LSTs. 
Based on the relative sensitivity tradeoff between $f/D$ and FoVs,
we find, for point-like sources, that optimal values are close to $f/D = 1.2$ and $\textrm{FoV} = 4.2^\circ$.
For extended sources ($\sim1^\circ$), the optimal values become larger: $f/D = 1.4$, $\textrm{FoV} = 4.5^\circ$.

\section{Observations under partial moonlight}

Observations under moonlight conditions permit to extend the observing time by as much 
as 30\%, based on experience with MAGIC \cite{Britzger2009} and VERITAS,
and are important for the detection of transient phenomena.
To estimate the performance of CTA under moonlight conditions,
MC simulations with a factor 4.5 higher night sky background (NSB) have been performed.
This corresponds to nights with the Moon above the horizon, illuminated at $\sim60\%$. 
In the core energy range around 1 TeV, analysis results indicate a similar 
performance to the one for dark sky conditions, in terms of sensitivity, angular 
resolution and energy resolution. However this is not the case for the low energy 
regime. Here, due to the higher noise levels, the energy thresholds are generally
a factor of 2 higher, although this number depends on the array layout.

\section{Readout simulations}

A number of tools exist to simulate the optical and electronic designs 
being evaluated within CTA (see \cite{Bernlohr2012} for a brief overview).
Here we outline ongoing studies that focus on the design of the camera readout.
One such study \cite{Vorobiov2011} aims to optimize the pixel integration time window,
which, given the short time scale of Cherenkov light flashes,
can minimize the error in collected pixel signal charge over NSB fluctuations.
For short integration times, the pixel charge resolution degrades due to Cherenkov light fluctuations,
while for longer times the NSB dominates.
A dynamic integration window, whose duration varies as a function of signal amplitude,
may improve charge resolution over a fixed duration window.
Under dark sky conditions, any improvements are expected to be very small,
as the Poissonian fluctuations of the Cherenkov signal dominate.
However under partial moonlight or Galactic plane observations more significant improvements can be envisaged.
A complementary study \cite{Naumann2011} proposes a flexible readout scheme in which 
only the pixels around the core of the shower image are recorded, with the aim of reduced 
data rates and effective dead times. The scheme would employ lower threshold local 
trigger information to define regions of interest in the camera and follow the shower 
temporal development, while a higher threshold trigger would trigger the camera readout.


\begin{theacknowledgments}
  We gratefully acknowledge support from the agencies and organizations
  listed in this page: \url{http://www.cta-observatory.org/?q=node/22}.
\end{theacknowledgments}

\end{document}